\documentclass[aps,twocolumn,showpacs]{revtex4}

\usepackage{bm}
\usepackage{epsfig}

\def\nb #1{{\hbox{\bf #1}}}

\newcommand{\lm}{{\cal P}}

\begin{document}

\title{Vortex Rings and Lieb Modes in a Cylindrical Bose-Einstein Condensate}
\author{S. Komineas$^1$ and N. Papanicolaou$^2$}
\affiliation{$^1$Physikalisches Institut, Universit\"at Bayreuth, 
D-95440 Bayreuth, Germany \\
$^2$Department of Physics, University of Crete, and Research Center of Crete, 
Heraklion, Greece \\
emails: stavros.komineas@uni-bayreuth.de \hspace{10pt} papanico@physics.uoc.gr}

\date{\today}

\begin{abstract}
We present a calculation of a solitary wave propagating along 
a cylindrical Bose-Einstein trap, which is found to be a hybrid of 
a one-dimensional (1D)
soliton and a three-dimensional (3D) vortex ring. The calculated 
energy-momentum dispersion exhibits characteristics similar to those of
a mode proposed sometime ago by Lieb within a 1D model, as well as some
rotonlike features.
\end{abstract}

\maketitle


Approximately forty years ago, Lieb and Liniger \cite{lieb1}
employed the Bethe Ansatz to obtain an exact diagonalization of the model
Hamiltonian that describes a one-dimensional (1D) Bose gas interacting via
a repulsive $\delta$-function potential. Based on this solution
Lieb \cite{lieb2} proposed an intriguing dual interpretation of the
spectrum of elementary excitations, either in terms of the familiar Bogoliubov
mode, or in terms of a certain kind of a particle-hole excitation
which will hereafter be referred to as the Lieb mode.
The energy-momentum dispersions of the Bogoliubov and the Lieb modes
coincide at low momenta, where they both display the linear dependence
characteristic of sound-wave propagation in an interacting Bose-Einstein
Condensate (BEC), but the Lieb dispersion is significantly different
at finite momenta where it exhibits a surprising $2\pi$ periodicity.
Yet this new mode remained somewhat of a theoretical curiosity
because of the absence of a physical realization of a strictly 1D Bose gas.

In a curious turn of events, the above subject re-emerged
in connection with the solitary waves calculated analytically
within a 1D classical Gross-Pitaevskii (GP) model \cite{tsuzuki,zakharov},
which were later shown to provide an approximate but fairly accurate
description of the quantum Lieb mode for practically
all coupling strengths \cite{kulish,ishikawa}. More importantly,
the same solitary waves motivated experimental observation
of similar coherent structures in BEC of alkali-metal atoms
through phase imprinting or phase engineering \cite{burger,denschlag}.
Therefore, an opportunity presents itself for experimental realization
of the Lieb mode.

It is clear that a theoretical investigation based on effective 1D models
often used to describe cylindrical traps will lead to the prediction
of a Lieb mode, by arguments similar to those employed in the
original 1D classical model \cite{kulish,ishikawa}. However,
one should also question the stability of the corresponding solitons
within the proper 3D environment of a realistic trap \cite{muryshev,feder}.
For example, dark solitons created within a spherical trap were
observed to decay into vortex rings \cite{anderson}.
This observation was also supported by a numerical calculation in
which an initial dark-soliton configuration is let to evolve in time
according to the GP equation.

Therefore, although the production of solitary waves through phase
imprinting clearly suggests some distinct 1D characteristics, it should
be expected that a proper understanding of such waves must also account
for 3D effects that are inevitably present in a realistic BEC.
One could envisage a picture in which the actual solitary wave is a hybrid
of a 1D soliton and a 3D vortex ring. It is the aim of the present
paper to make the above claim precise by calculating solitons propagating
along a cylindrical trap without making {\it a priori} assumptions about their
effective dimensionality. Our approach was motivated by the calculation
of vortex rings in a homogeneous BEC due to Jones and Roberts \cite{jones1}
and a similar calculation of  semitopological solitons in
planar ferromagnets \cite{semi}.

Explicit results will be presented for a set of parameters that roughly
correspond to the experiment of Ref. \cite{burger}.
Thus we consider a cigar-shaped trap filled with $^{87}$Rb atoms,
with a transverse confinement frequency 
$\omega_\bot \!=\! 2\pi\! \times\! 425$ Hz
and a corresponding oscillator length 
$a_\bot \!=\! \sqrt{\hbar/m\omega_\bot} \!\approx\! 0.5\mu m$.
The coupling constant is written as $U_o \!=\! 4\pi\hbar^2 a/m$
where $a\!\approx\! 50\, \hbox{\AA}$ is the scattering length. 
We make the approximation
of an infinitely-long cylindrical trap with average linear density
$\nu\!=\!0.2\, \hbox{atoms}/\hbox{\AA}$ and introduce the dimensionless
combinations of parameters $\gamma\!=\! \nu a\!=\! 10$ and
$\gamma_\bot\!=\! \nu a_\bot\!=\! 10^3$.
Finally, rationalized units are defined through the rescalings
$t\!\to\! t/\omega_\bot,\;\; {\nb r}\!\to\! a_\bot {\nb r},\;\; \hbox{and}\;\;
\Psi\!\to\! \sqrt{\nu}\, \Psi/ a_\bot\,$.

The energy functional extended to include a chemical potential is then
given by
\begin{equation}
\label{eq:energy}
  W = \frac{1}{2} \int{\left[ \bm{\nabla}\Psi^* \bm{\nabla}\Psi
 + \rho^2\,\Psi^*\Psi + g (\Psi^*\Psi)^2 - 2\mu\,\Psi^*\Psi \right]\, dV},
\end{equation}
where $g\!=\!4\pi\gamma$ and $\rho^2\!=\!x^2+y^2$, and yields energy in units
of $\gamma_\bot\hbar\omega_\bot$ while the chemical potential $\mu$
is still measured in units of $\hbar \omega_\bot$. Similarly, the
conserved linear momentum along the $z$ axis given by the usual definition
\begin{equation}
\label{eq:momentum}
\lm = \frac{1}{2 i}\, \int{\left(\Psi^* \frac{\partial\Psi}{\partial z}
-\frac{\partial\Psi^*}{\partial z}\Psi\right)\; dV} = 
\int{n \frac{\partial\phi}{\partial z}\; dV},
\end{equation}
is measured in units of $\hbar \nu\!=\!\gamma_\bot (\hbar/a_\bot)$.
In the second step of Eq.~(\ref{eq:momentum}) we employed the usual 
hydrodynamic variables defined from $\Psi\!=\!\sqrt{n}\, e^{ i\phi}$.

\begin{figure}
   \begin{center}
   \epsfig{file=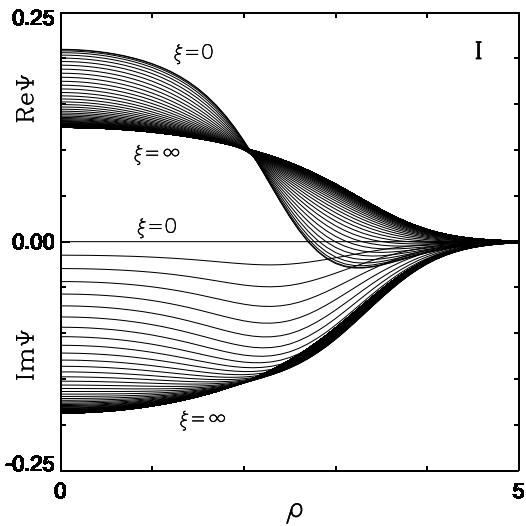,width=7cm,bbllx=140bp,bblly=430bp,bburx=430bp,bbury=720bp}
   \end{center}
   \caption{Radial dependence of the solitary wave function at
$v\!=\!c/2$ for various positive values of $\xi$ in steps of 
$\delta\xi\!=\!0.1$. The result for negative $\xi$ is obtained through the
symmetry relations $\hbox{Re}\,\Psi(\rho,\xi)\!=\!\hbox{Re}\,\Psi(\rho,-\xi)$
and $\hbox{Im}\,\Psi(\rho,\xi)\!=\!-\hbox{Im}\,\Psi(\rho,-\xi)$.
Distances are measured in units of $a_\bot$.
   }
   \label{fig1}
\end{figure} 

An important first step in out calculation was to obtain accurate 
information about the ground state and the corresponding linear (Bogoliubov)
modes. The ground-state wave function $\Psi\!=\!\Psi_0(\rho)$
was calculated as a stationary point of the energy functional
(\ref{eq:energy}) by a variant of a relaxation algorithm
\cite{dalfovo} and is normalized according to 
$\int_0^\infty{2\pi\rho\, d\rho\; |\Psi_0|^2}\!=\!1$
to conform with our choice of rationalized units.
The chemical potential was found to be $\mu\!=\!6.4324$ for $\gamma\!=\!10$.
We have also recalculated the lowest branch of the Bogoliubov spectrum
from which we extracted the speed of sound $c\!=\! 1.77$ in units
of $a_\bot \omega_\bot$. This numerical estimate at $\gamma\!=\!10$
is consistent to within 1\% with the Thomas-Fermi approximation
$c\!=\!\gamma^{1/4}$ which was previously derived in a number of papers
\cite{zaremba,kavoulakis,stringari} and was employed for the
analysis of experimental data \cite{andrews}.

Thus we turn to the calculation of axially-symmetric solitary waves
described by a wave function of the form
$\Psi\!=\!\Psi(\rho,\xi)$ where $\xi\!=\!z-vt$ and $v$ is the 
constant velocity along the $z$ axis. Such a wave 
function satisfies the stationary differential equation
\begin{eqnarray}
\label{eq:equation}
 - i v\, \frac{\partial\Psi}{\partial\xi} & = & -\frac{1}{2} \Delta\Psi
 + \frac{1}{2} \rho^2\Psi + g (\Psi^*\Psi) \Psi - \mu \Psi\,, \\
\noalign{\medskip}
 \Delta & = & \frac{\partial^2}{\partial\rho^2} + \frac{1}{\rho}
 \frac{\partial}{\partial\rho} + \frac{\partial^2}{\partial\xi^2}\,,
\nonumber
\end{eqnarray}
which is supplemented by the boundary conditions that $\Psi$ vanish for
$\rho\!\to\!\infty$, and $|\Psi|\!\to\!|\Psi_0(\rho)|$ 
for $\xi\!\to\!\pm\infty$.
The phase of the wave function is not fixed {\it a priori} at spatial infinity,
except for a mild restriction implied by the Neumann boundary
condition $\partial\Psi/\partial\xi\!=\! 0$ imposed at
$\xi\to\pm\infty$ mainly for numerical purposes. Finally, a solution of
Eq.~(\ref{eq:equation}) must satisfy the virial relation
\begin{equation}
\label{eq:virial}
v\, \lm = \int{\left[\frac{1}{2} \frac{\partial\Psi^*}{\partial\xi}
\frac{\partial\Psi}{\partial\xi} + \rho^2\,\Psi^*\Psi
+ \frac{g}{2} (\Psi^*\Psi)^2 - \mu \Psi^*\Psi \right] dV}\,,
\end{equation}
obtained by a standard scaling argument.

\begin{figure}
   \begin{center}
   \epsfig{file=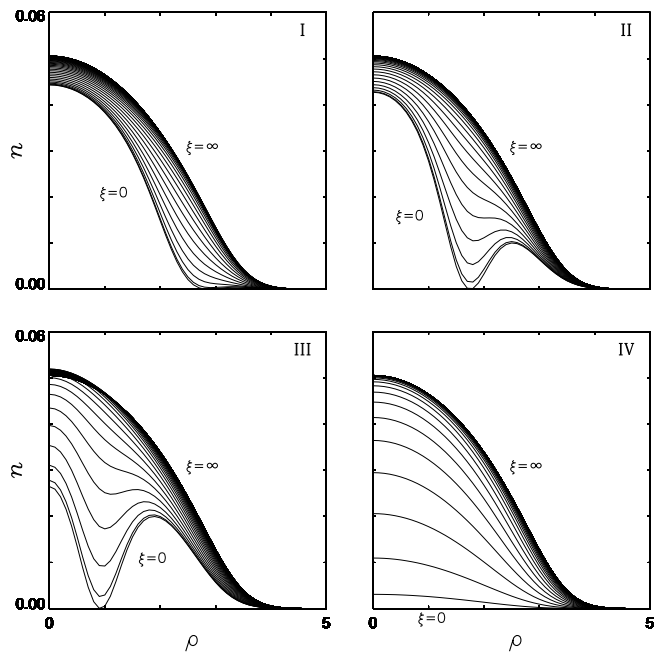,width=8.5cm,bbllx=40bp,bblly=190bp,bburx=545bp,bbury=705bp}
   \end{center}
   \caption{Radial dependence of the local particle density $n\!=\!|\Psi|^2$
for the four special cases discussed in the text. Conventions are the 
same as in Fig.~\ref{fig1}.
   }
   \label{fig2}
\end{figure}

Solutions of Eq.~(\ref{eq:equation}) were obtained by a sophisticated
Newton-Raphson iterative algorithm \cite{jones1,semi} which will not
be described here in any detail. For values of the velocity $v$
near the speed of sound $c$, the calculated solitary wave is a rarefaction
pulse that may be thought of as a mild soundlike disturbance of the
ground state. The dominant features of the solitary wave are pronounced
as the velocity is decreased to lower values and become reasonably
apparent at $v\!=\!c/2$, as shown in Fig.~\ref{fig1} which provides a complete
illustration of the wave function through its real and imaginary parts.
A partial but more transparent illustration is given in frame I of
Fig.~\ref{fig2} where we depict the radial dependence of the local
particle density $n\!=\!|\Psi|^2$ for various values of $\xi$.
It is clear that the density near the center of the soliton ($\xi\!=\!0$)
vanishes on a ring with a relatively large radius $R\!\simeq\! 2.8$,
thus a vortex ring is beginning to emerge.
The features of the vortex ring become completely apparent,
and its radius is tightened, as we proceed to yet smaller values 
of the velocity.
A notable special case is the static ($v\!=\!0$) vortex ring with radius
$R\!\simeq\! 1.8$ illustrated in frame II of Fig.~\ref{fig2}, which
is far from being a completely dark (black) solitary wave.

One would think that pushing the velocity $v$ to negative values 
would retrace the calculated sequence of solitary waves backwards.
In fact, our algorithm continues to converge to vortex rings
of smaller radii until a critical velocity $v\!=\!-v_0\!\simeq\!-0.475\, c$
is encountered where the vortex ring achieves its minimum radius
$R_{\rm min}\!\simeq\! 0.8$ and ceases to exist for smaller values of
$v$. The terminal state at $v\!=\!-v_0$ is illustrated in frame III
of Fig.~\ref{fig2}. We have thus described a sequence of solitary waves
with velocities in the range $-v_0\!<\!v\!<\!c$, which does not 
contain a black soliton.
An equivalent sequence is obtained in the range $-c\!<\!v\!<\!v_0$
simply by reversing the relative sign between the real and imaginary
part of the wave function, as is evident from Eq.~(\ref{eq:equation}).

It is now important to calculate the energy-momentum dispersion.
The excitation energy is defined as $E\!=\!W-W_0$ where both $W$ and $W_0$
are calculated from Eq.~(\ref{eq:energy}) applied  for the solitary wave
$\Psi(\rho,\xi)$ and the ground state $\Psi_0(\rho)$, respectively.
The presence of the chemical potential in Eq.~(\ref{eq:energy})
provides the compensation that is necessary in order to compare
energies of states with the same number of particles.
Similarly, the relevant ``physical'' momentum is not the linear
momentum $\lm$ of Eq.~(\ref{eq:momentum}) but the ``impulse'' $Q$
defined in a manner analogous to the case of a homogeneous gas
\cite{ishikawa,jones1}:
\begin{eqnarray}
\label{eq:impulse}
 Q  & = & \int{(n-n_0) \frac{\partial\phi}{\partial z}\, dV} 
          = \lm - \delta\phi \\
\noalign{\medskip}
 \delta \phi & \equiv & \int_0^\infty{2\pi\rho\, d\rho\; n_0(\rho)
[\phi(\rho,z\!=\! \infty)-\phi(\rho,z\!=\! -\infty)]}\,, \nonumber
\end{eqnarray}
where $n_0\!=\!|\Psi_0(\rho)|^2$ is the ground-state particle density
and $\delta \phi$ is the weighted average of the phase difference
between the two ends of the trap. Here we simply postulate the validity
of definition (\ref{eq:impulse}) and note that the corresponding
group-velocity relation $v\!=\!dE/dQ$ was satisfied to an excellent
accuracy in our numerical calculation. On the other hand, the virial
relation (\ref{eq:virial}) was verified using the standard linear momentum
$\lm$ of Eq.~(\ref{eq:momentum}), as expected.

\begin{figure}
   \begin{center}
   \epsfig{file=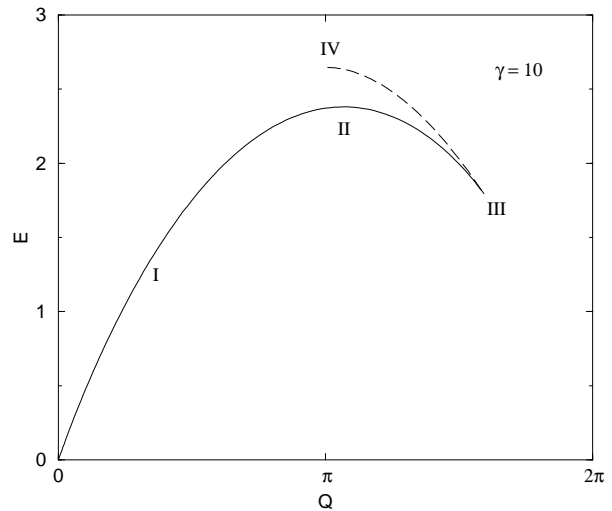,width=8cm}
   \end{center}
   \caption{Energy $E$ in units of $\gamma_\bot (\hbar\omega_\bot)$ versus
impulse $Q$ in units of $\gamma_\bot (\hbar/a_\bot)$. The solid line
corresponds to the main sequence of solitary waves discussed in the text,
and the dashed line to the auxiliary sequence that contains a black
soliton (point IV).
   }
   \label{fig3}
\end{figure}

The calculated dispersion $E\!=\!E(Q)$ is depicted in Fig.~\ref{fig3}
by a solid line along which we place the symbols I, II and III
that correspond to the special cases of the solitary wave
discussed above. At low $Q$ the dispersion is linear, $E\!=\!c\,|Q|$,
a feature that it shares with the Bogoliubov dispersion. However,
the energy now reaches a maximum at point II where $v\!=\! 0$ and $Q$
is slightly greater that $\pi$. Interestingly, the group velocity  becomes
negative in the region (II,III) or, equivalently, the impulse is
opposite to the group velocity. This rotonlike behavior could develop
into a full-scale roton if the terminal point III actually turns out to
be an inflection point beyond which the group velocity may begin to rise again.
We have not been able to somehow continue our sequence of solitary waves
beyond point III, nor have we ruled out such a possibility.
In any case, the picture just described, together with the fact 
that the calculated radius of the vortex ring is monotonically
decreasing along the sequence (I,II,III) comes close to the Onsager-Feynman
view of a roton as the ghost of a vanished vortex ring \cite{donnelly}.

The branch of the dispersion shown by a solid line
in Fig.~\ref{fig3} was calculated on the basis of the sequence
of solitary waves with velocities in the range $-v_0 \!<\! v \!<\! c$.
The branch for $-c \!<\! v \!<\! v_0$ may be thought to correspond to
negative $Q$, as usual. However, there is an implicit $2\pi$
periodicity because of the appearance of the angle $\phi$
in the definition of the impulse in Eq.~(\ref{eq:impulse}).
Therefore, a more natural representation of the spectrum seems to be
obtained by restricting $Q$ to the ``first Brillouin zone''
and thus completing Fig.~\ref{fig3} by its mirror image around $Q\!=\!\pi$.

We have also carried out a detailed numerical calculation of the 
Bogoliubov dispersion for comparison. But the main point can be made
here by assuming the model dispersion
$\omega=|q| (c^2+q^2/4)^{1/2}$ where the frequency $\omega$ is measured
in units of $\omega_\bot$ and the wave number $q$ in units of $1/a_\bot$.
Translated into the units employed in Fig.~\ref{fig3}, the model
Bogoliubov dispersion reads $E=|Q| (c^2+ \gamma_\bot^2 Q^2/4)^{1/2}$
where a huge factor $\gamma_\bot^2\!\approx\! 10^6$ makes its appearance.
Therefore, although the Bogoliubov and Lieb dispersions coincide at low
$Q$, they diverge quickly at the scale of Fig.~\ref{fig3},
a notable difference from the homogeneous 1D model \cite{lieb2,ishikawa}.
In other words, the two modes operate at rather different energy
and momentum scales in a cylindrical trap.

The last question we address is whether or not there exists an
independent sequence of solitary waves that reduces to a black soliton
at $v\!=\!0$. We used a model black-soliton configuration as input
in our iterative algorithm which converged to the true
static soliton illustrated in frame IV of Fig.~\ref{fig2} that is
indeed a black soliton. We then incremented the velocity to either
positive or negative values with symmetrical results.
For definiteness, we follow the solution to negative $v$ and find that
it exists only until one encounters the same critical velocity
$-v_0$ discussed earlier in the text. The corresponding terminal state
is precisely the same with the one reached through our original
sequence of solitary waves and illustrated in frame III of Fig.~\ref{fig2}.
For intermediate values of $v$ the solution is again a vortex ring
with {\it constant} radius at $R\!=\!R_{\rm min}\!\approx\!0.8$.
Accordingly, the calculated energy-momentum dispersion shown by
a dashed line in Fig.~\ref{fig3} joins the original sequence through a cusp
at the terminal point III, a situation that is reminiscent of the
calculation of Jones and Roberts \cite{jones1}.
The same authors together with Putterman later argued \cite{jones2}
that the solitary waves that correspond to the upper branch are
actually unstable, a conclusion that might be valid in the present
case as well.

\begin{figure}
   \begin{center}
   \epsfig{file=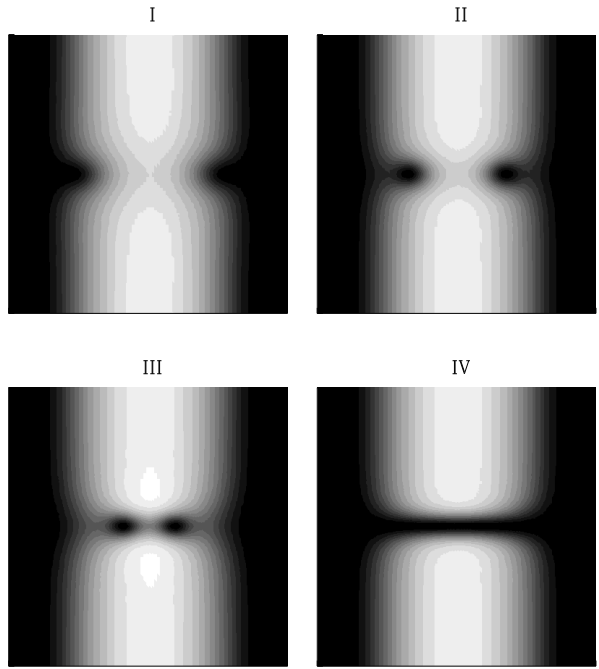,width=8cm,bbllx=20bp,bblly=130bp,bburx=565bp,bbury=755bp}
   \end{center}
   \caption{Contour levels of the local particle density $n$ in a plane that
contains the $z$ axis and cuts across the cylindrical trap. The complete
3D picture may be envisaged by simple revolution around the $z$ axis.
Regions with high particle density are bright while regions with zero density
are black. The four special cases considered are the same as in
Fig.~\ref{fig2}.
   }
   \label{fig4}
\end{figure}

A summary view of our results is given in Fig.~\ref{fig4}. We have
thus constructed a family of solitary waves which exhibit some quasi-1D
features, such as the appearance of a nontrivial phase difference
$\delta\phi$ that is important for phase engineering \cite{burger,denschlag},
but are otherwise bonafide 3D vortex rings.
\vspace{1cm}

We are grateful to G.M. Kavoulakis for numerous discussions
that guided us through the extensive recent literature on BEC.
S.K. acknowledges financial support from the Gratuiertenkolleg
``Non-equilibrium phenomena and phase transitions in complex systems''
and thanks the University of Crete and the Research Center of Crete
for hospitality.


\end{document}